\documentstyle[11pt,paspconf,psfig]{article}

\def\apj{ApJ}
\def\aap{A\&A}

\def\nDutchString{Thirtynine} %% = 36 + 3 comuni
\def\nJKT{16}                 %% = 13 + 3 comuni

\def\commonDutchJKT{Three}   %% questi ammassi osservati sia al Sud sia al Nord
\def\nTwoMeter{16}           %% ca. 16 GC al INT fino m-M = 18 

\def\mMUnMetro{16}           %% m-M di quelli osservati con telescopi 1 m
\def\mMDuMetri{18}           %% m-M di quelli osservati con telescopi 2 m

\def\ZPerror{0.03}           %% errore max nel zero point delle calibrazioni

     \def\sigmaSC{0.1} %% metodo verticale: rms dei residui attorno a SCL97
\def\vsigVIstrip{0.01 } %% V98 entro +/- 2 Gyr
\def\rgcmax{22}    %% Fino a quale Rgc si spingono le nostre osservazioni?

\def\aVdb98{0.046}
\def\bVdb98{0.41}
\def\cVdb98{1.15}

       \def\hbA{0.20} % 
       \def\hbB{0.98} %
    \def\hbBv98{0.90} % come sopra, per vdb98

\def\vertv98{2}
\def\verts97{5}
\def\orizv98{3}
\def\orizs97{6}
\def\meanv98{4}
\def\means97{7}

\def\sigmaVs97{0.10}
\def\sigmats97{1.2}
\def\sigmaVv98{0.11}
\def\sigmatv98{1.3}
\def\sigmacoevverts97{0.59}
\def\sigmacoevvertv98{0.76}

\def\sigmaviv98{0.018}
\def\sigmavis97{0.015}
\def\sigmatviv98{1.3}
\def\sigmatvis97{1.1}
\def\sigmacoevorizs97{}  %% non e' prevista dalla macro, ma del resto
\def\sigmacoevorizv98{}  %% il campione orizzontale praticamente coincide
\begin{document}
\title{Relative ages of inner-halo globular clusters}

\author{Ivo Saviane}
\affil{Dipartimento di Astronomia, Universit\`a Padova}
\author{Alfred Rosenberg}
\affil{Telescopio Nazionale Galileo, Osservatorio Astronomico Padova}
\author{Giampaolo Piotto}
\affil{Dipartimento di Astronomia, Universit\`a Padova}

\begin{abstract}

We present preliminary results from our on-going survey of Galactic
globular clusters relative ages. The investigation is based on $V,I$
images obtained at the ESO and La Palma telescopes, which make up the
largest homogenous photometric catalog to date. A second, independent
sample of globulars observed in the $B, V$ bands provides an
independent check of the results based on the groundbased data.

Age-dependent morphological parameters are measured on the CMDs and
are compared with two sets of independent models. We find that the
so-called ``vertical'' and ``horizontal'' methods give compatible
results when the groundbased dataset is used, and that in both cases
the observed trends are well reproduced by the isochrones. The
interpretation of the {\sc hst} data trends are more controversial,
due to both a stronger dependence on metallicity, and to the discrepancies
in the theoretical loci.

Our data clearly show that (a) no age dispersion can be revealed for
the bulk of the GGCs at the $\sim 1$~Gyr level; (b) no age-metallicity
relation is found, although the age dispersion is somwhat larger for
intermediate and higher metallicity clusters; and (c) there is no clear
trend with the galactocentric distance, out to the
present limits of our survey ($R_{\rm GC} < \rgcmax$ Kpc).

\end{abstract}


\section{Introduction} \label{sec:intro}

Galactic globular clusters (GGC) are the oldest components of the 
Galactic halo. The determination of their relative ages and of any age
correlation with metallicities, abundance patterns, positions and
kinematics allows to establish the formation timescale of the halo and
gives information on the early efficiency of the enrichment processes
in the proto--galactic material. The importance of these problems and
the difficulty in answering to these questions is at the basis of
the huge efforts dedicated to gather the relative ages of GGCs in
the last 30 years or so (VandenBerg, Stetson, and Bolte 1996,
Sarajedini, Chaboyer, Demarque 1997, SCD97, and references therein).

Any method for the age determination of GGCs is based on the position
of the turnoff (TO) in the color--magnitude diagram (CMD) of
their stellar population. We can measure either the absolute
magnitude or the de--reddened color of the TO. In order to overcome
the uncertainties intrinsic to any method to get GGCs distances and
reddening, it is common to measure either the color or the magnitude
(or both!) of the TO, relative to some other point in the CMD whose
position does not depend on age.

%%%%%%%%%%%%%%%%%%%%%%%%%%%%%%%%%%%%%%%%%%%%%%%%%%%% S T A R T   T E X T

%%%%%%%%%%%%%%%%%%%%%%%%%%%%%%%%%%%%%%%%%%%%%%%%%%%% S T A R T   T E X T

Observationally, as pointed out by Sarajedini \& Demarque (1990)
and VandenBerg et al. (1990, VBS90), the most precise relative age
indicator is based on the TO color relative to some fixed point on the
red giant branch (RGB).  Unfortunately, the theoretical RGB
temperature is very sensitive to the adopted mixing length parameter,
whose dependence on the metallicity is not established yet. As a
consequence, investigations on relative ages based on this method
(``horizontal method'') might be of difficult interpretation, and need
a careful calibration of the relative TO color as a function of the
relative age (Buonanno et al. 1998, B98).  The other age indicator
is based on the TO luminosity relative to the horizontal branch
(HB). Though this is usually considered a more robust relative age
indicator, it is affected both by the uncertainty on the dependence
of the HB luminosity on metallicity and the empirical
difficulties to get both the TO magnitude and the HB magnitude for
clusters with only blue HBs.

Despite the intrinsic difficulties in gathering relative ages, it is
nevertheless astonishing, for 
those not working in the field,
to read
the totally contradictory  results coming from different groups.

We are still debating whether GGCs are almost coeval (Stetson et al. 1996)
or whether the GGCs have continued to form for 5 Gyr (SCD97) or so 
(i.e. for 30-40\% of the Galactic halo lifetime).

Indeed, there is a major limitation to the large scale GGC relative age
investigations: the photometric inhomogeneity and the inhomogeneity in
the analysis of the databases used in the various studies. And even
worst, these etherogeneous collections of data do not allow a
reliable treatment of the empirical errors, which sometimes must
be guessed, with questionable results (Chaboyer et al. 1996).

Prompted by this major drawback, two years ago our group began the
collection of an homogeneous photometric material for a large sample of
GGCs, in order to obtain accurate relative ages by using both the
horizontal and vertical method in a self-consistent way.  The strategy was
decided after a preliminary analysis of published CMDs both in the $B,V$
and $V,I$ bands (Saviane, Rosenberg, and Piotto 1997; hereafter
SRP97). SRP97 showed that the $V-I$ color differences are less sensitive to
metallicity than the $B-V$ ones (while retaining the same age
sensitivity). SRP97 also suggested that a high-precision, large-scale
investigation in the $V$ and $I$ bands
would have allowed a relative age determination through the horizontal
method without the usual limitation of dividing the clusters into
different metallicity groups (VBS90).

Here we present the first exciting results of this investigation.

\section{Data base} \label{sec:database}

In the present investigation only two telescopes (one for the northern
and one for the southern sky GGCs) have been used.

\nDutchString\ clusters have been observed with the ESO/Dutch 0.9m telescope at La
Silla, and \nJKT\ at the RGO/JKT 1m telescope in la Palma. A total of
30 clusters had CMDs useful for the relative age determinations.

In this observing campaign (the first step of our investigation)
all the clusters with $(m-M)_V<\mMUnMetro$ have been observed with 1-m class
telescopes. We have also observed \nTwoMeter\ clusters within
$(m-M)_V<\mMDuMetri$ with 2-m class telescopes, and observations at 4-m class
telescopes for the farthest clusters are planned.

The data have been calibrated with the same set of standards.
The observations, reduction, and photometry will be described in
forthcoming papers.  Here suffice to say that the
zero-point uncertainties of our calibrations are $<\ZPerror$~mag for each
band. \commonDutchJKT\ clusters were observed both with the southern and the
northern telescopes, thus providing a consistency check of the
calibrations: no systematic differences were found, at the level of
accuracy of the zero-points.

We are also collecting an independent and even more homogeneous
database in the $B$, and $V$ bands. The data come from two {\sc hst}
programs (GO6095 and GO7470). Within GO7470 we should observe with the
{\sc WFPC2} the core of 46 clusters; With the already available
archive data by the end of GO7470, all the GGCs with $(m-M)_B<18$
should have been observed with HST.  Though the programme main
objectives are different, most of the data are suitable for this
project.  This database allows an independent check of the results
from the groundbased data.

In order to have well defined fiducial lines for each CMD, a selection
on the photometric catalogs of each cluster was applied by imposing a
threshold on the photometric errors, and only the less crowded regions
were used.  The following points were then measured on the CMD, both
for the {\sc HST} and groundbased samples: Magnitude and color of the TO;
Magnitude of the MS point 0.05~mag redder than the TO; Color of the
RGB at $\Delta m$ magnitudes above one of the two previous points
(where $\Delta m$ was 1.5, 2.0, 2.5, 3.0, 3.5); The magnitude level of
the HB.  These values were used to calculate a set of both vertical
and horizontal parameters.  We will name these parameters,
generically, $\delta x_{@y}$ or $\delta x_{@y}^{0.05}$. For example,
$\delta (V-I)_{@1.5}^{0.05}$ is the difference between the $(V-I)$
color of the RGB and that of the TO. In this case, the RGB point is
measured 1.5~mag above the MS point 0.05~mag redder than the TO.
In the following, we will use $\Delta V_{\rm HB}^{\rm TO}$ as vertical
parameter and $\delta (V-I)_{@2.5}$ as horizontal parameter. However,
the results presented below are independent from this choice, as
will be shown in Rosenberg, Saviane, and Piotto (1999, RSP99).

\section{Methodology} \label{sec:analisi}

Basically, we followed the B98 strategy. In view of the uncertainties
associated to the interpretation of the horizontal parameter (cf. Section
\ref{sec:intro}), we first identified a set of coeval clusters by means of
the vertical method. These coeval GGCs allowed to identify an empirical
``isochrone'' in the $\delta$~color vs. [Fe/H] plane (a straight line in
B98). These isochrones were then compared with the theoretical predictions.
Finally, the color differences from the mean line were converted into an age.

The choice of the metallicity scale will be discussed in details in
RSP99. In view of its homogeneity, we used the Rutledge et al.
(1997) compilation on the  Carretta \& Gratton
(1997) metallicity scale.

\subsection{Coeval clusters}

\begin{figure}[t]
\vspace{0cm}

\hbox{
\hspace{0cm}

\psfig{figure=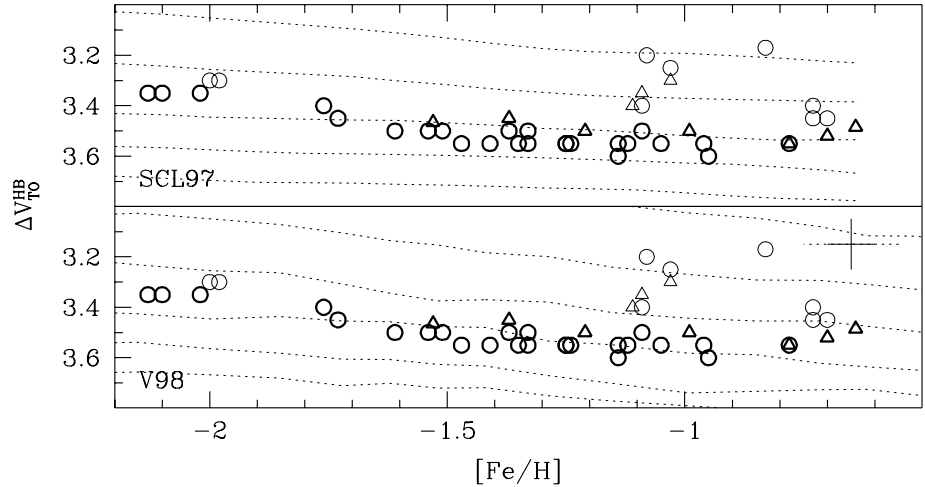,width=4.8in}
}

\vspace{0cm}

\caption[]{
The parameter $\Delta V_{\rm HB}^{\rm TO}$ is plotted versus the
metallicity. The dashed lines in the two panels show the theoretical trend
of the parameter for the models of SCL97 (top) and V98 (bottom), assuming
that $M_V^{\rm HB}=\hbA {\rm [Fe/H]} + \hbB$.
The isocrones are separated by 2 Gyr.  Open circles identify the
groundbased sample, while open triangles identify the {\sc HST}
sample; the typical error is shown by the cross in the lower panel.
Heavy symbols represent the fiducial coeval clusters. selected as
described in the text.
}
\label{fig:voverplot}
\end{figure}

\begin{figure}[t]
\vspace{0cm}

\hbox{
\hspace{0cm}

\psfig{figure=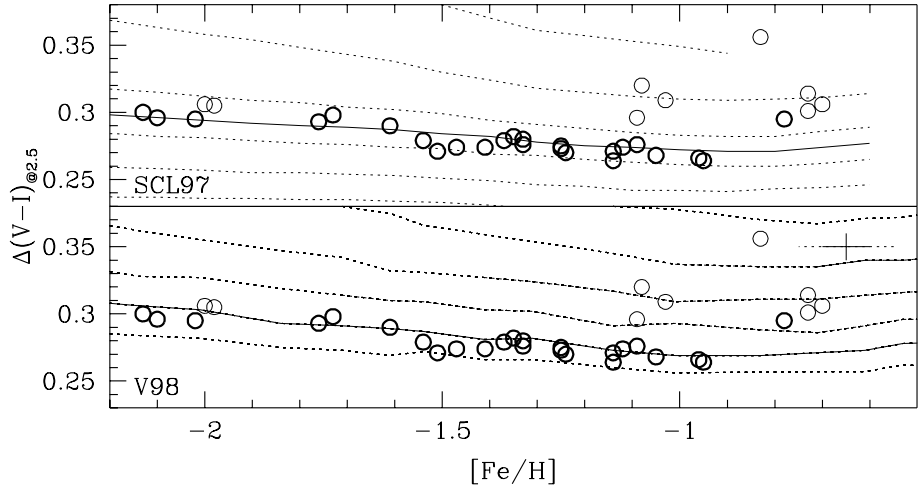,width=4.8in} 
}

\vspace{0cm}

\caption[]{
Plot of the parameter $\delta (V-I)_{@2.5}$ vs. metallicity for the groundbased
sample. Two sets of theoretical models
are also represented (dashed lines), SCL97 (top panel) and V98 (bottom
panel). Again, heavy symbols mark the fiducial coeval clusters,
and the typical
errors are represented by the cross. The isochrone used for the relative
age determination is displayed as a solid line
 }
\label{fig:h_vi_25}
\end{figure}

\begin{figure}[t]
\vspace{0cm}

\hbox{
\hspace{0cm}

\psfig{figure=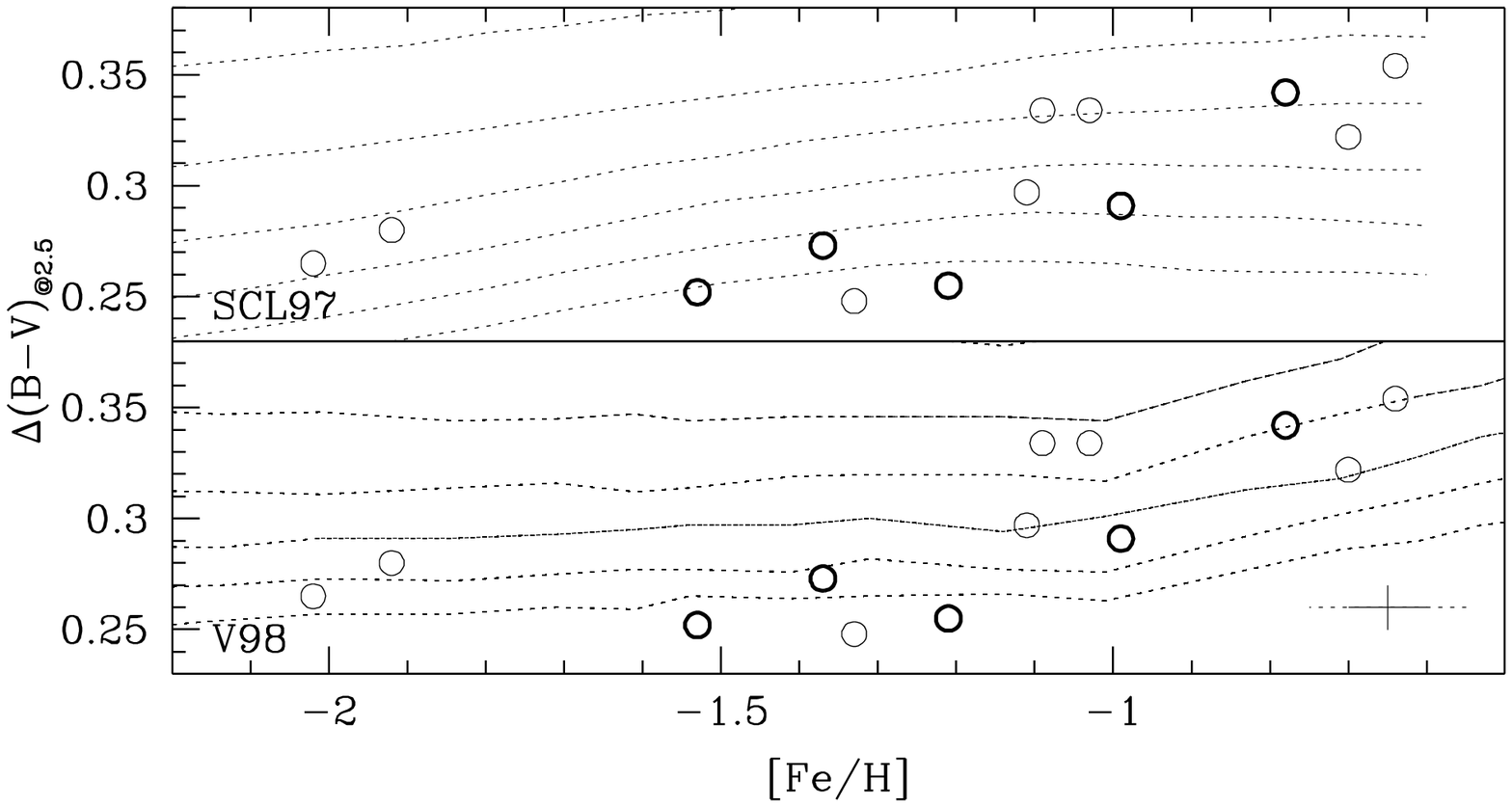,width=4.8in} 
}

\vspace{0cm}

\caption[]{
Same as Fig.~2 for  the parameter $\delta (B-V)$. No relative ages have
been computed from this sample, so all isochrones are represented as dashed
lines

}
\label{fig:h_bv_25}
\end{figure}

In Fig.~\ref{fig:voverplot}, the parameter $\Delta V_{\rm HB}^{\rm
TO}$ is plotted vs.  metallicity, both for the groundbased and the
{\sc hst} sample of GGCs. In the same figure, the theoretical
isochrones are represented as dashed lines. The theoretical $\Delta
V_{\rm HB}^{\rm TO}$ was calculated from the TO of VandenBerg et
al. (1998, V98) and Straniero et al. (1997) models, and assuming
$V_{\rm HB} = \hbA \cdot {\rm [Fe/H]} + \hbB$ (Chaboyer et al. 1996).
These models were chosen, since they are the most recent ones offering
both $B-V$ and $V-I$ colors.

With our choice for the $V_{\rm HB}$ vs. [Fe/H] relation, and within
the observational errors, the
theoretical isochrones and the observed values show similar trends
with metallicity. 
It must be clearly stated that this
result depends on the choice of the theoretical HB luminosity, though
the conclusions would be the same if the slope of the $V_{\rm HB}$
vs. [Fe/H] relation is changed by not more than $\pm 15\%$(see also below).  
Note that
the zero point of the relation for  $V_{\rm HB}$ does not affect the
relative age.  The isochrones can be used to tentatively select a
sample of coeval clusters. We will use these clusters to test
the isochrones in the $\delta (V-I)_{@2.5}$ vs [Fe/H] plane
(B98).
We somehow arbitrarely defined as coeval (from here on
fiducial coeval GGCs), those clusters whose vertical parameter was
within $\pm 1\,\sigma$ from the isochrone which better fit the data
distribution in the $V_{\rm HB}$ vs.[Fe/H] plane.  
These object are marked by heavy symbols in Fig.~\ref{fig:voverplot}. 
Interestingly enough, the same set of coeval clusters
is selected using either the SCL97 or the V98 isochrones, and using a
slope $\alpha$ for the $V_{\rm HB}$ vs. [Fe/H] relation in the range
$0.17<\alpha<0.23$ for the V98 isochrones and $0.15<\alpha<0.20$ for the
SCL97 isochrones.
The observed dispersion is $\sigma = \sigmaSC$~mag 
with respect to both the SCL97 and V98 isochrones, 
i.e. fully compatible with the uncertainties in $\Delta
V_{\rm HB}^{\rm TO}$, strengthening the idea that the selected
clusters must be coeval.

\subsection{Ages from color differences} \label{sec:orizage}

In Fig.~\ref{fig:h_vi_25}, the parameter $\delta (V-I)_{@2.5}$
vs. metallicity for the groundbased sample is compared with the SCL97 ({\it
top panel}) and V98 ({\it bottom panel}) isochrones.  The trend with
metallicity of the $\delta (V-I)_{@2.5}$ parameter for the fiducial coeval
GGCs ({\it filled circles}) is remarkably similar to the theoretical
trend. In Fig.~\ref{fig:h_vi_25}, the fiducial coeval GGCs are all within a
2~Gyr strip, showing a full consistency with what was found from the
vertical method.

The plot in Fig.~\ref{fig:h_bv_25} is the $B-V$ counterpart of
Fig.~\ref{fig:h_vi_25}. Also in this case most of the GGCs are within a
narrow band. However, as pointed out also by B98, the age width of this
band is more difficult to obtain, since the isochorones show different
trends with [Fe/H]. Also the trend with [Fe/H] of the $\delta (B-V)_{@2.5}$
for the coeval clusters differs from the isochrones.  The differences in
$\delta (B-V)_{@2.5}$ for different models and different bolometric
corrections are widely discussed in B98. Here, we simply note that the
recent V98 calculations seem to better approximate the observed data and
that, using these isochrones, an age dispersion comparable with that from
the vertical method is obtained.

A further remark on the different dependence of the horizontal parameters
in $(B-V)$ and in $(V-I)$ on the metallicity.  Fig.~\ref{fig:h_vi_25} and
\ref{fig:h_bv_25} are plotted on the same scale. Clearly, $\delta
(B-V)_{@2.5}$ strongly depends on [Fe/H], particularly for [F/H]$\geq
-1.7$, as already pointed out by VBS90. As a consequence, even a small
error on the metal content of a cluster can strongly affect the
determination of its relative age. This fact might also explain the
apparently larger dispersion of the $\delta (B-V)_{@2.5}$ parameter.
$\delta (V-I)_{@2.5}$ has a much milder dependence on metallicity.

All the above considerations strengthen the conclusions by SRP97 that the
$\delta (V-I)$ parameter is much more reliable than the $\delta (B-V)$ as a
relative age index.

Relative ages were computed only by means of the difference in the
$\delta (V-I)_{@2.5}$ parameter with respect to the 13~Gyr-SCL97 or
14~Gyr-V98 isochrone fitted to the points.  The $\delta (V-I)_{@2.5}$
dispersion is
\vsigVIstrip~mag, as expected on the basis of the errors in measuring this
parameter.

\section{Discussion}

\begin{figure}[t]
\vspace{0cm}

\hbox{
\hspace{0cm}

\begin{tabular}{c}
\psfig{figure=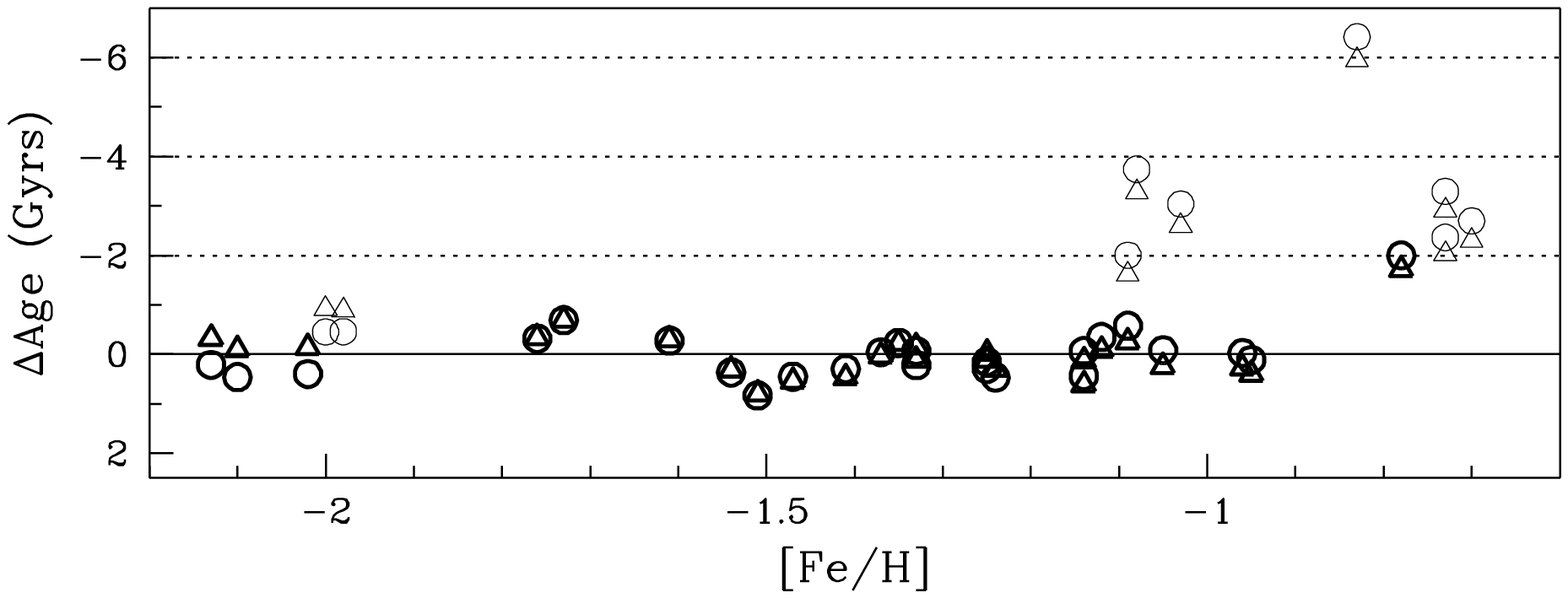,width=4.8in} \\
\psfig{figure=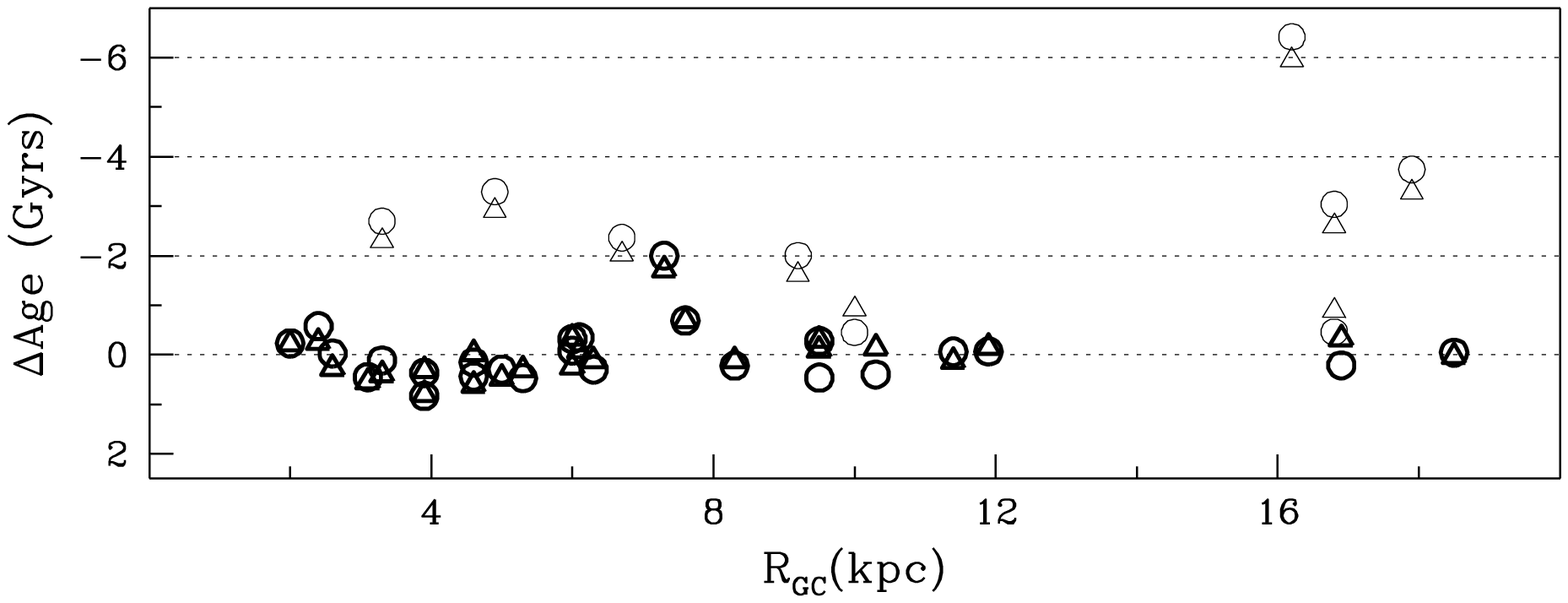,width=4.8in}
\end{tabular}
}

\vspace{0cm}

\caption[]{
Relative ages obtained from the observed $\delta (V-I)$  using  two
theoretical relations (see text for details). The trends vs. [Fe/H] (top
panel) and the Galactocentric radius $R_{\rm GC}$ (bottom panel) are
shown. Open circles represent values obtained using the V98 isochrones,
while open triangles are the values obtained using those of SCL97
}
\label{fig:age_feh_rgc}
\end{figure}

The dispersions in $\delta (V-I)_{@2.5}$ 
translate in an age dispersion of 1.4~Gyr
 (adopting the SCL97 models) or
1.6~Gyr 
(adopting the V98 models), which lowers to 1.3 and 1.4 Gyr if we a remove 
Pal~12, a known anomalously young cluster (Rosenberg et al. 1998).
The age dispersion of the adopted coeval clusters is of 0.75 Gyr for the
SCL97 models and 0.70 for the V98 models.

As pointed out above, if we take into account the observational errors, the
GGC age dispersion is {\bf fully compatible with a null age dispersion}.

The relative ages from the horizontal method estimated from
Fig.~\ref{fig:voverplot} and Fig.~\ref{fig:h_vi_25} are plotted in
Fig.~\ref{fig:age_feh_rgc} vs. [Fe/H] and the Galactocentric distance
$R_{GC}$. The {\it open circles} are the ages from V98 models and the
{\it open triangles} represents the ages from the SCL97 models.
Regardless of the model, the relative ages do not depend on the
cluster metallicity, though the age dispersion is larger for the
intermediate and higher metallicity GGCs. No clear dependence on the
galactocentric distance can be identified.

These results indicate that the bulk of the Galactic halo formed on a
timescale $\le 1$~Gyr; a minor fraction of younger clusters is also
present, although their true Galactic origin is still debated. These
younger clusters tend to be located in the outer halo: the
interpretation of this trend is controversial. They could have 
formed in isolated Searle \& Zinn (1978) fragments later
accreted into the halo, or else they could be explained by the SGMC GC
model formation of Harris \& Pudritz (1994). In this context a
delayed formation of the outer GCs is naturally explained (see also
Harris et al. 1998).

An age-metallicity relation cannot be detected by this
investigation. This means that the early chemical enrichment of the
Galactic halo took place on a timescale again $< 1$~Gyr, up to values
roughly half solar.

\acknowledgments
It is a pleasure to thank Peter Stetson for his generosity in providing
us with all the software needed for the stellar photometry and for the
helpful discussions. Alessandro Chieffi and Don VandenBerg are charmly
thanked for the discussions and suggestions and for making available
their models in advance of publication.

\end{document}